\documentclass[11pt,letterpaper]{article}

\usepackage[margin=2.8cm]{geometry}

\usepackage{amsfonts}
\usepackage{amssymb}
\usepackage{mathrsfs}
\usepackage{amsmath,amssymb}
\usepackage{subeqn}
\usepackage{bbm}

\makeatletter\renewcommand{\section}{\@startsection
{section}{1}{\z@}{-2.5ex plus -1ex minus
    -.2ex}{2.3ex plus .2ex}{\centering\large\sc }}

\makeatletter\renewcommand{\subsection}{\@startsection{subsection}{2}{\z@}{-3.25ex
plus -1ex minus
   -.2ex}{1.5ex plus .2ex}{\centering\large\sc }}

\makeatletter\renewcommand{\subsubsection}{\@startsection{subsubsection}{3}{-2.45ex}{-3.25ex
plus -1ex minus -.2ex}{1.5ex plus .2ex}{\it }}

\makeatletter\renewcommand{\paragraph}{\@startsection{paragraph}{4}{\z@}%
                                    {0.8ex \@plus1ex \@minus.2ex}%
                                    {-.5em}%
                                    {\normalfont\normalsize\bfseries}}

\renewcommand{\thesection}{\arabic{section}.}

\numberwithin{paragraph}{section}
\renewcommand\theparagraph {\S\thesection\@arabic\c@paragraph.\kern-8pt}

\numberwithin{equation}{section}

\renewcommand*\l@section{\@dottedtocline{1}{0em}{1.5em}}
\renewcommand*\l@subsubsection{\@dottedtocline{4}{3.8em}{3.2em}}

\renewcommand\tableofcontents{%
    \section*{\large\contentsname
        \@mkboth{%
           \MakeUppercase\contentsname}{\MakeUppercase\contentsname}}%
       {\baselineskip=15pt plus 2pt minus 1pt
    \@starttoc{toc}}%
\vspace{-3mm}\centerline{{\vrule height 0.5pt width 15.5cm depth
0pt}} }

\renewenvironment{thebibliography}[1]
     {\section*{\centering{\refname}
        \@mkboth{\MakeUppercase\refname}{\MakeUppercase\refname}}%
     \list{\@biblabel{\@arabic\c@enumiv}}%
           {\settowidth\labelwidth{\@biblabel{#1}}%
            \leftmargin\labelwidth
            \advance\leftmargin\labelsep
            \@openbib@code
            \usecounter{enumiv}%
            \let\p@enumiv\@empty
            \renewcommand\theenumiv{\@arabic\c@enumiv}}%
      \sloppy
      \clubpenalty4000
      \@clubpenalty \clubpenalty
      \widowpenalty4000%
      \sfcode`\.\@m}

\DeclareFontFamily{U}{rsf}{}
\DeclareFontShape{U}{rsf}{m}{n}{
  <5> <6> rsfs5 <7> <8> <9> rsfs7 <10-> rsfs10}{}
\DeclareMathAlphabet\Scr{U}{rsf}{m}{n}

\newcommand{\p}{\partial}

\renewcommand{\d}{\mathrm{d}}
\newcommand{\im}{\mathrm{i}}

\newcommand{\IC}{\mathbbm{C}}
\newcommand{\IR}{\mathbbm{R}}
\newcommand{\IZ}{\mathbbm{Z}}

\newcommand{\cC}{\mathscr{C}}

\newcommand{\CL}{\mathcal{L}}
\newcommand{\CN}{\mathcal{N}}

\newcommand{\al}{{\alpha}}
\newcommand{\be}{{\beta}}

\def \bi{\bibitem}
\def \bs {\bigskip}
\def \la{\label}
\def \ci{\cite}

\def \td {\tilde}
\newcommand{\rf}[1]{(\ref{#1})}

\begin{document}

\begin{titlepage}

\setcounter{page}{0}
\renewcommand{\thefootnote}{\fnsymbol{footnote}}

\begin{flushright}
 Imperial--TP--RR--04/2007
\end{flushright}

\vspace*{1cm}

\begin{center}

{\LARGE\bf On T-Duality and Integrability for Strings}\\[3pt]
{\bf\LARGE on AdS Backgrounds}

\vspace*{1cm}

{\Large
 Riccardo Ricci$^{a,b}$, Arkady A. Tseytlin$^{a,c,}$\footnote{Also at
 the Lebedev Institute, Moscow, Russia.}
 and Martin Wolf$^{a,c}$ \footnote{{\it E-mail addresses:\/} 
 {\ttfamily r.ricci, a.tseytlin, m.wolf@imperial.ac.uk}}
}\\

\vspace*{1cm}

{\it $^{a}$ Theoretical Physics Group\\ 
The Blackett Laboratory, Imperial College London\\
Prince Consort Road, London SW7 2AZ, United Kingdom}\\[.5cm]

{\it $^{b}$ The Institute for Mathematical Sciences\\
Imperial College London\\
Prince's Gate, London SW7 2PG, United Kingdom}\\[.5cm]

{\it $^{c}$ The Isaac Newton Institute for Mathematical Sciences\\
20 Clarkson Road, Cambridge CB3 0EH, United Kingdom}

\vspace*{1cm}

{\bf Abstract}

\end{center}

\begin{quote}
We discuss an  interplay between T-duality and
integrability for certain  classical non-linear sigma models.
In particular, we consider strings on the AdS$_5\times S^5$ background
and perform T-duality along the four isometry directions of AdS$_5$
in the Poincar\'e patch. The T-dual of the
AdS$_5$ sigma model  is again a sigma model on an 
AdS$_5$ space. This classical  T-duality relation 
was used in the recently uncovered  connection
between  light-like Wilson loops and MHV  gluon scattering amplitudes 
in the  strong coupling limit of  the  AdS/CFT duality. 
We show that the explicit coordinate dependence along the T-duality
directions of the associated Lax connection (flat current) can be eliminated 
by means of a field dependent gauge transformation. As a result, the gauge
equivalent Lax connection can easily be
T-dualized, i.e.~written in terms of the dual set
of isometric coordinates. The T-dual Lax connection can be used
for the derivation of infinitely many
conserved charges in the T-dual model. Our construction implies that 
 local (Noether) charges of  the original model are mapped to
non-local charges of  the T-dual model and vice versa.

\vfill\noindent November 5, 2007

\end{quote}

\setcounter{footnote}{0}\renewcommand{\thefootnote}{\arabic{thefootnote}}

\end{titlepage}

\section{\kern-10pt Introduction and summary}

Some of the recent remarkable  advances in our understanding 
of $\CN=4$ superconformal Yang-Mills 
theory have been possible due to its  integrability
in the planar limit. This integrability was observed 
in the dilatation  operator (or the spectrum of anomalous dimensions)
 at several  leading orders in a weak coupling  expansion and is  expected to hold 
to all orders (see, e.g.,  \cite{Beisert:2004ry} and references therein).
Integrability is present  also in the  strong coupling 
limit  of the $\CN=4$ SYM 
 theory as described  by the AdS$_5\times S^5$  superstring \cite{Metsaev:1998it}.
Indeed, the well-known  classical 
integrability  of the bosonic AdS$_5\times S^5$ sigma model   was shown in \cite{Bena:2003wd}
to extend to its $\kappa$-symmetric Green-Schwarz-type fermionic generalization
\cite{Metsaev:1998it}. In particular, an infinite number of
conserved non-local charges for the classical superstring was found
(see also \ci{kmmz}).\footnote{Aspects related to involutivity of the
charges  were discussed in  \cite{Dorey:2006mx}.
 The existence of such charges
was also verified in the pure spinor formulation of the AdS$_5\times S^5$ 
 superstring \cite{Vallilo:2003nx} and shown to persist
after quantization
\cite{Berkovits:2004jw,Mikhailov:2007mr}.} 
This   integrability  appears to extend also to one-  and two-loop orders in the 
quantum string sigma model  as  implied   by the 
computations  of   quantum  string corrections to semi-classical  string energies
and their matching  to predictions of the interpolating Bethe ansatz
(see \ci{bes,ort,rt} and references there). 
The  non-local conserved  charges found on the 
string side  appear to have a counterpart
in planar gauge theory at weak coupling within the
spin-chain formulation   for   the dilatation operator
  \cite{Dolan:2003uh}, but their direct 
interpretation in a field theoretic language is still
missing.\footnote{See also \cite{Wolf:2004hp} for recent developments about the twistor
approach to non-local symmetries.}

\bs

One expects that the integrability of the  $\CN=4$ SYM 
 theory  should not only have  important consequences for the spectrum 
 of anomalous  dimensions  of gauge-invariant single trace operators (i.e.~for 
the spectrum of energies of the {\it closed}  AdS$_5\times S^5$ superstring)  but also 
 for other  observables, e.g., 
 for the structure of expectation  values of certain Wilson
  loops.
 The dual AdS/CFT counterpart of
 the latter are partition functions of {\it open} AdS$_5\times S^5$ strings ending 
 on  certain contours at the boundary of AdS$_5$ \ci{malrey}.
 The integrability can be used, e.g., for finding the corresponding minimal
  surfaces.\footnote{See 
 \cite{moreloops} for some  applications of 
 integrability to  the computation of   a wide class of Wilson loops.}
 With a motivation to 
 study  quantum string corrections to
  Wilson loops, Ref.~\ci{kt} considered  a special  $\kappa$-symmetry 
 gauge \ci{Kallosh:1998qv} in which the action  of \cite{Metsaev:1998it}
 written in the Poincar\'e coordinates of AdS$_5$
  simplifies to 
\begin{equation}
 \begin{aligned}
 S[X,Y,\Theta]\ &=\ - \frac{T}{2}\int_\Sigma \left[{Y^2} \,\eta_{ab}(\d X^a  
 - 2 \im \bar{\Theta}\Gamma^a \d\Theta)\wedge{*(\d X^b  
 - 2 \im \bar{\Theta}\Gamma^b \d\Theta)}\
 + \right. \\
 & \kern2cm\left. +\ \frac{1}{Y^2}\delta_{ij} \d Y^i\wedge{*\d Y^j}
+ 4\im \,\delta_{ij}\d Y^i \bar{\Theta}\wedge  \Gamma^j \d\Theta\right]\!. \la{acc} 
\end{aligned}
\end{equation}
Here, $T=\frac{\sqrt\lambda}{2 \pi}$ is the string tension, 
 `$*$' is the Hodge star on the Minkowski world-sheet $\Sigma$, 
$X^a$ are  the four coordinates in the directions  
parallel  to the boundary of 
AdS$_5$  with $(\eta_{ab})={\rm diag}(1,1,1,-1)$ ($a,b=1,\ldots,4$) and  $Y^i$ are the six remaining 
coordinates with $Y^2= \delta_{ij}Y^iY^j$ and $(\delta_{ij})={\rm diag}(1,\ldots,1)$  ($i,j=5,\ldots,10$). Furthermore, $\Theta$ 
is a  ten-dimensional Majorana-Weyl spinor and $\Gamma=(\Gamma^a,\Gamma^i)$ are the standard 
``flat'' ten-dimensional Dirac matrices. 
 It was observed in \ci{kt} that since the  action \eqref{acc} depends on the isometric coordinates 
 $X^a$ only through their  derivatives, it  
 can be  simplified further  if one trades  $X^a$ for the  set  of four dual 
  two-dimensional scalars $\tilde X^a$, i.e.~if one
 performs the formal  T-duality transformation \ci{bush,roc}  along $X^a$. 
 Remarkably, the bosonic  part of the resulting  action 
 has again an AdS$_5\times S^5$ geometry (with $Y^2 \mapsto {1/Y^2}, 
 \ X^a \mapsto \tilde X^a$) 
  and its fermionic part  becomes  simply  quadratic in $\Theta$ 
  \begin{equation}
\begin{aligned}
\tilde S[\tilde X,Y,\Theta]\ &=\ - \frac{T}{2} \int_\Sigma \left[ \ \frac{1}{Y^2}
\left(  \eta_{ab}\d\tilde X^a\wedge{*\d\tilde X^b}+  
 \delta_{ij} \d Y^i\wedge{*\d Y^j}\right)
\ +\right.\\
&\kern2.5cm\left.+\ 4\im \,\bar{\Theta}\left(\eta_{ab}\d\tilde X^a \Gamma^b+ 
\delta_{ij}\d Y^i \Gamma^j\right)\wedge \d\Theta\right]\! . \la{bcc} 
\end{aligned}
\end{equation}
Like  the  bosonic part of the original action \rf{acc}, the  bosonic part of the 
T-dual action \rf{bcc} 
has an $SO(4,2) \times SO(6)$  global symmetry,  with  the two $SO(4,2)$ conformal groups acting,
 of course,  on  different 
(dual)  sets of  variables.\footnote{The full actions have only 
$SO(3,1) \times SO(6)$ as an obvious linearly realized symmetry. 
They are also invariant under the  scaling  transformations  
($X^a \mapsto \ell  X^a, \   Y \mapsto  \ell^{-1}   Y$ and
 $\td X^a \mapsto \ell^{-1} \td X^a$) (the invariance of the fermionic term 
 can be seen  by writing $Y^i = Y n^i, \  n^i n_i =1$
and  rescaling the fermions  by $Y^{1/2}$-factor using that for Majorana-Weyl spinors 
$\bar{\Theta}\Gamma\Theta =0$). Other global symmetries of \rf{acc}  are broken by the choice of 
the $\kappa$-symmetry gauge (i.e.~they are present modulo a gauge transformation).
Note also  that  quantum 
  T-duality transformation  produces also a dilaton term
 $\Phi = - 2 \ln (Y^2)$  which  formally breaks the ``dual'' conformal symmetry.
  Here, we shall ignore it as 
 all considerations in this paper will  be classical.}
 
Since the $X^a$-directions  are non-compact, this  T-duality is not an equivalence transformation 
on a two-dimensional cylinder, i.e.~the   transformed action is  not 
  appropriate as a starting point  for the study of
 the closed  string  spectrum of the  AdS$_5\times S^5$ superstring. However,  
  it  may still  be useful in the {\it open} string context
  which we will have in mind.

Indeed, this T-dual
formulation appears to play an
important  role in the recently discovered  connection between 
maximally helicity-violating
(MHV) gluon scattering amplitudes \ci{bern}
and special  Wilson loops (defined  on contours 
 formed  by light-like gluon momentum vectors) 
 at strong \cite{am1,am3} \footnote{The T-dual action 
\eqref{bcc} was used also  for quantum one- and two-loop 
string computations in this context 
in \ci{krtt,rt}.} and weak  \ci{dr,br}  coupling.
 The classical $SO(4,2)$ conformal symmetry of
the T-dual AdS geometry seems to have something to do  with  the mysterious 
 ``dual'' conformal symmetry observed in the momentum-space integrands 
 of loop integrals  for planar gluon scattering amplitudes 
  \ci{dr,dc}.
From the point of
view of the original AdS$_5\times S^5$  model, this dual conformal symmetry could
be related to the presence of hidden symmetries associated with the 
integrability and, as such, it may correspond to
the existence of non-local currents.

\bs

With a  motivation  to  shed some light on these issues, it is
 essential to understand how the 
integrable structure emerges in the T-dual formulation and, 
 in
particular, how it translates from the original model to the
T-dual one. Since T-duality   maps classical solutions
 to classical solutions and also 
since  the T-dual geometry is again 
 AdS$_5 \times S^5$,  it is reasonable to expect
that the T-dual model  is also  integrable. 
One question then is how to
map the Lax connection of the original model to the T-dual one, 
or  how  to map the non-local flat currents and the
associated non-local conserved charges to the T-dual
model.  Some previous  work on T-duality in the context of related integrable models 
appeared in  \ci{Frolov:2005dj,hat,kluson:2007} (see also \ci{Arutyunov:2005nk,old}).

\bs

Below  we shall perform a first step in this direction by 
focusing for simplicity on the bosonic part of the model \rf{acc} and \rf{bcc}.
 After having presented the general
setting in the next section, we shall then  discuss the T-duality for two
toy examples: the two-sphere $S^2$ and the two-dimensional anti-de
Sitter space AdS$_2$. These two examples are   different in
nature as T-duality will be performed along a compact direction
for $S^2$ and along a non-compact direction for AdS$_2$. On the
other hand, they both exhibit some generic features
useful to understand in view of our eventual aim --  the sigma
model on AdS$_5\times S^5$. 

To perform the T-duality explicitly,
 one needs to express the T-dual coordinates in terms of the original 
ones. This relation is non-local;  that  makes it  non-trivial 
to solve for the T-dual coordinates and to find 
the  T-duality image  of flat currents. 
 Nevertheless,
 we demonstrate that it is possible to
eliminate  the explicit dependence of the flat currents 
on the  T-duality direction 
coordinate     by means of a finite field dependent 
gauge
transformation. This yields a gauge equivalent Lax connection that
depends only on the derivatives of the isometric coordinates. 
Our procedure is  similar to the one in 
 \cite{Frolov:2005dj} (see also \cite{kluson:2007}). 
 We
then go on to  discuss T-duality for general AdS geometries
including AdS$_5$ case.

\bs 

Having gauged away the explicit coordinate dependence, the 
T-duality on the flat currents of the AdS$_5$ 
sigma model  can be easily implemented. This in
turn  allows us to find the 
T-dual flat currents representing the  dual $\widetilde{SO(4,2)}$ conformal symmetry. 
An application of this formalism would be the explicit 
construction of an  infinite tower of 
 conserved charges
 in both the original and the T-dual 
AdS$_5$ spaces and  the investigation   of their relation. It is likely, that these
charges will only be well defined after a suitable regularization
(cf.~\ci{am1}).\footnote{We thank F. Alday  for drawing our attention 
to this  issue.} 

\bs 

A natural next step   would  be an  extension of
the present analysis  to the full superstring action \rf{acc}.
While the  
T-duality acts on the bosonic AdS coordinates  in a relatively simple way, so that, 
e.g., the full conformal symmetry group  present before
  the duality reappears after it, 
the situation is not as simple when one includes the 
fermions. The two dual  actions \rf{acc} and \rf{bcc} do  not appear to be  related 
 by a local  change of variables and renaming  the fields
  (as it was  at the bosonic level).
 In particular,  some of the $PSU(2,2|4)$ superisometries 
 present in \rf{acc} (those preserved by the $\kappa$-symmetry gauge fixing) 
are not  manifest in the T-dual formalism.
It would  be important to understand whether
the original symmetry group is still present but realized in some hidden,
non-local way, as it is in the example  of  a much simpler 
bosonic sigma model on $S^2$ discussed in Sec.~\ref{sec:twosphere}
This may turn out to be  helpful  for 
a better understanding of Wilson loops in the
T-dual AdS$_5$ space \cite{am1,am3}. 

\bs

\section{\kern-10pt  Non-linear sigma models and non-local charges}\label{sec:general}

\subsection{\kern-10pt Generalities}\label{sec:gen}

\paragraph{Non-linear sigma models.}
Let us consider the non-linear sigma model action
\begin{equation}\label{eq:sma}
S[X]\ =\ -\tfrac{1}{2}\int_\Sigma g_{IJ}(X)\, \d X^I\wedge{*\d} X^J,
\end{equation}
where $X:(\Sigma,h)\to(M,g)$ embeds a  pseudo-Riemannian surface $\Sigma$
with metric $h$ into some
$d$-dimensional  pseudo-Riemannian manifold $M$ (target space)
with metric $g$; $I,J,\ldots=0,\ldots,d-1$.
Below, we shall coordinatize $\Sigma$ by $t$ and $x$.
Furthermore, we assume that there is some connected Lie group $G$ acting
on $M$ by isometries. This implies that for any $V\in\mathfrak{g}:={\rm Lie}(G)$ we
have a corresponding vector field $\xi_V$,
\begin{equation}
\xi_V(f)\ :=\ \left.\frac{\d}{\d t}\right|_{t=0} f\circ\exp(tV),
\end{equation}
for some function $f$ on $M$,
such that \eqref{eq:sma} is invariant under
\begin{equation}
X^I\ \mapsto\ X^I+\xi^I_V,\qquad\mbox{with}\qquad
\CL_{\xi_V} g\ =\ 0.
\end{equation}
Here, $\CL_{\xi_V}$ denotes the Lie derivative along
the vector field $\xi_V$. Put differently, for any $V\in\mathfrak{g}$,
the vector field $\xi_V$ is a Killing vector field of $g$.

As a short calculation reveals,
the Noether current associated with the Killing vector field $\xi_V$
takes the form\footnote{Note that this
current does not coincide with the one obtained from the
stress tensor but differs by an improvement term.}
\begin{equation}\label{eq:currents}
\langle j,V\rangle\ =\ g_{IJ}(X)\d X^I\xi^J_V(X),
\end{equation}
where $\langle\cdot,\cdot\rangle$ is a metric
on $\mathfrak{g}$.

\paragraph{Flat currents and charges.}
Suppose now that in addition to being conserved, $\d{*j}=0$, the
current $j$ also
satisfies a flatness condition
\begin{equation}\la{fla}
\d j+j\wedge j\ =\ 0.
\end{equation}
This occurs, for instance, when $M$ is a symmetric space $G/H$,
where $G$ is the isometry group of $M$ and $H$ is the isotropy group
of the action of $G$ on $M$ on some fixed $p\in M$. More generally,
such flatness conditions arise when $M$ is a coset space that admits a
certain $\IZ_m$-grading \cite{Young:2005jv}.

Given some conserved current $j$ which
is flat, then  there
are always two one-parameter families of flat currents $J$ (Lax connection).
Indeed, by considering general linear combinations
of the form (with $a,b$ being  real  numbers) 
\begin{equation}\label{eq:zc}
J\ =\ a\,j+b\,{*j},
\end{equation}
one observes that\footnote{Here we used
${**}1=1$ and ${*\al}\wedge\be+\al\wedge{*\be}=0$,
for $\al,\be\in\Omega^1\Sigma$.
}
\begin{equation}
\d J+J\wedge J\ =\ (a^2-a-b^2)j\wedge j\ \overset{!}{=}\ 0.
\end{equation}
This can be solved by putting
\begin{equation}\la{las}
 J\ =\ J_{\lambda\pm}\ :\qquad a\ =\ \tfrac{1}{2}[1\pm\cosh (\lambda) ]\qquad\mbox{and}
\qquad b\ =\ \tfrac{1}{2}\sinh (\lambda)\ ,
\end{equation}
for $\lambda\in\IR$. Note that in that case
the zero
curvature equation \eqref{eq:zc} encodes the equations of motion
for the action \eqref{eq:sma}.

As was shown in 
\cite{Luscher:1977rq}, 
taking the path-ordered exponential
\begin{equation}
\label{pathorder}
W(t,x;t_0,x_0|J_{\lambda\pm})\ =\ P\exp\left(-\int_\cC J_{\lambda\pm}\right)\!
\end{equation}
along some unbounded spatial contour $\cC\subset\Sigma$ at
some fixed time, a given
one-parameter family of flat currents $J_{\lambda\pm}$
always induces an
infinite number of conserved non-local charges. To be
concrete, for $J_{\lambda-}$, the quantity
\begin{equation} Q_{\lambda-}(t)\ =\ \lim_{x\to\infty}\, W(t,x;t,-x|J_{\lambda-})
       \ =\ 1+\sum_{n=1}^\infty \lambda^n Q_n(t)\label{charges}
\end{equation}
is conserved for all $\lambda$, i.e.
\begin{equation} \frac{\d}{\d t} Q_n(t)\ =\ 0,\end{equation}
 provided $j$ has an
appropriate fall-off at spatial infinity. The
first two charges read as
\begin{equation}
\begin{aligned}
 Q_1(t)\ &=\ \tfrac{1}{2}\int_{-\infty}^\infty\d x\, j_0(t,x),\\
 Q_2(t)\ &=\ -\tfrac{1}{4}\int_{-\infty}^\infty\d x\, j_1(t,x)+
             \tfrac{1}{2}\int_{-\infty}^\infty\d x \int_{-\infty}^x\d x'\,
                         j_0(t,x)j_0(t,x').
\end{aligned}
\end{equation}
The charge $Q_2(t)$ generates,  via Poisson brackets,  all the higher charges
$Q_n(t)$  (for more details, see, e.g., the review in  \cite{Dolan:1983bp}).

\subsection{\kern-10pt AdS geometries}

Let us now specify the case  of AdS geometries that  we shall 
be discussing later.  In particular, we  shall derive
the Killing vectors and the Noether currents in the Poincar\'e-patch 
 parametrization  we are interested in. 

\paragraph{Setting.}
Consider the sigma model on the direct product $M={\rm
AdS}_p\times S^{d-p}$.
We coordinatize $M$
by $(X^a,Y^i)$ with $a,b,\ldots=1,\ldots,p-1$ and
$i,j,\ldots=p,\ldots,d$, respectively, and equip it with the
(conformally flat) metric
\begin{equation}\la{ini}
g\ =\ \tfrac{1}{Y^2}(\eta_{ab}\,\d X^a\otimes\d X^b+\delta_{ij}\,
               \d Y^i\otimes\d Y^j).
\end{equation}
Here, we have introduced the following abbreviations:
$Y^2:=\delta_{ij}Y^i Y^j$, $(\delta_{ij})={\rm diag}(1,\ldots,1)$
and $(\eta_{ab})$ $={\rm diag} (\underbrace{1,\ldots,1, }_{p-r-1}
 \ \underbrace{\ -1,\ldots,-1}_{r})
$, where $r=0$ for the Euclidean   and $r=1$ for the Minkowski AdS 
spaces.

Note that the metric $g$ can be brought into its standard form
by performing the following change of coordinates:
\begin{subequations}
\begin{equation}
(X^a,Y^i)\ \mapsto\ ( X^a,\hat Y^i)\ :=\ (X^a,Y^{-1} Y^i).
\end{equation}
In these coordinates, $g$ reads as
\begin{equation}
g\ =\ \underbrace{\tfrac{1}{Y^2}(\eta_{ab}\,\d X^a\otimes\d X^b+
          \d Y\otimes\d Y)}_{{\rm AdS}_p{\rm -part}}+
          \underbrace{\delta_{ij}\, \d\hat Y^i\otimes\d\hat Y^j}_{S^{d-p}{\rm -part}},
\end{equation}
\end{subequations}
with $\delta_{ij}\hat Y^i\hat Y^j=1$.

Below, when discussing T-duality, we shall choose to start with 
the AdS metric given in \rf{ini}, 
i.e.~we will do T-duality in the ``opposite'' direction compared to \rf{acc} and \rf{bcc}.
This is, of course, simply a convention as the two  choices are related
by
\begin{subequations}
\begin{equation} (X^a,Y^i)\ \mapsto\ ( X^a,\bar Y^i)\ :=\ (X^a, Y^{-2} Y^i)
\end{equation}
and which results in
\begin{equation} g\ =\ \eta_{ab}\,\bar Y^2\d X^a\otimes\d  X^b+
        \tfrac{1}{\bar Y^2}  \delta_{ij}\, 
       \d\bar Y^i\otimes\d\bar Y^j,
\end{equation}
\end{subequations}
with $\bar Y^2:=\delta_{ij}\bar Y^i\bar Y^j$. The choice \rf{ini} will help us
to simplify notation.

\paragraph{Killing vectors.}
Recalling the coset representations
\begin{equation}
{\rm AdS}_p\ \cong\ SO(p-r,r+1)/SO(p-r,r)
   \quad\mbox{and}\quad
   S^q\ \cong\ SO(q+1)/SO(q),
\end{equation}
we observe that the isometry group of $M$ is $G\cong
SO(p-r,r+1)\times SO(d-p+1)\subset SO(d,r+1)$. Hence, there
are $\frac{1}{2}d(d+1)-p(d-p)$ Killing vectors which represent
$\mathfrak{g}\cong\mathfrak{so}(p-r,r+1)\oplus\mathfrak{so}(d-p+1)$.
In the $(X^a,Y^j)$ parametrization
of $M$, they are given by
\begin{equation}
\begin{aligned}
  \xi_{L_{ab}} \ &=\ X_a\p_b-X_b\p_a,\qquad
  \xi_{M_{ij}}\ =\ Y_i\p_j-Y_j\p_i,\\
  \xi_{P_a} \ &=\ \p_a,\qquad
  \xi_D\ =\ X^a\p_a+Y^i\p_i,\\
  \xi_{K_a} \ &=\ (X^2+Y^2)\p_a-2X_a(X^b\p_b+Y^i\p_i),
\end{aligned}
\end{equation}
where $X_a:=\eta_{ab}X^b$ and $Y_i:=\delta_{ij}Y^j$, $\p_a:=\p/\p
X^a$ and $\p_i:=\p/\p Y^i$ and $X^2:=\eta_{ab}X^aX^b=X_aX^a$.
 The $\mathfrak{so}(p-r,r+1)$ and
$\mathfrak{so}(d-p+1)$ Lie algebras are generated by
$L_{ab}, P_a, D, K_a$ and $M_{ij}$, respectively.

The non-vanishing commutators among the above vector fields are:
\begin{equation}
\begin{aligned}
  {[\xi_{L_{ab}},\xi_{L_{cd}}]} \ &=\
  \eta_{bc}\xi_{L_{ad}}-\eta_{bd}\xi_{L_{ac}}-
  \eta_{ac}\xi_{L_{bd}}+\eta_{ad}\xi_{L_{bc}},\\
  {[\xi_{L_{ab}},\xi_{P_c}]}  \ &=\
    \eta_{bc}\xi_{P_a}-\eta_{ac}\xi_{P_b},\qquad
   {[\xi_{L_{ab}},\xi_{K_c}]} \ =\
    \eta_{bc}\xi_{K_a}-\eta_{ac}\xi_{K_b},\\
   {[\xi_{P_a},\xi_D]}  \ &=\  \xi_{P_a},\qquad
   {[\xi_{K_a},\xi_D]} \ =\ -\xi_{K_a},\\
   {[\xi_{P_a},\xi_{K_b}]}  \ &=\  2\xi_{L_{ab}}
     -2\eta_{ab}\xi_D,\\
 {[\xi_{M_{ij}},\xi_{M_{kl}}]}  \ &=\
  \eta_{jk}\xi_{M_{il}}-\eta_{jl}\xi_{M_{ik}}-
  \eta_{ik}\xi_{M_{jl}}+\eta_{il}\xi_{M_{jk}}.
\end{aligned}
\end{equation}

\paragraph{Flat currents.}
Knowing all the Killing vectors fields, we are now in the
position to construct the
associated Noether currents by using the formula \eqref{eq:currents}.
In the present situation, it reads as
\begin{equation}
 \langle j,V\rangle\ =\ \tfrac{1}{Y^2}(\eta_{ab}\d X^a\xi^b_V+
       \delta_{ij}\d Y^i\xi^j_V),
\end{equation}
where $\xi_V$ represents any of the above Killing vectors fields
for $V\in\mathfrak{g}$.
Using their explicit expressions, we obtain
\begin{equation}
\begin{aligned}
  \langle j, L_{ab}\rangle  \ &=\  \tfrac{1}{Y^2}
                 (\d X_aX_b-\d X_bX_a),\qquad
  \langle j, M_{ij}\rangle \ =\  -\tfrac{1}{Y^2}
                 (\d Y_iY_j-\d Y_jY_i),\\
  \langle j, P_a\rangle   \ &=\   \tfrac{1}{Y^2}\d X_a,\qquad
  \langle j, D\rangle\ =\ -\tfrac{2}{Y^2}(\d X^a X_a+\d Y^i Y_i),\\
  \langle j, K_a\rangle \ &=\  \tfrac{1}{Y^2}(X^2+Y^2)\d X_a-
           \tfrac{2}{Y^2}X_a(\d X^bX_b+\d Y^iY_i).
\end{aligned}
\end{equation}

Then the current $j$, as constructed
above, satisfies a flatness condition $\d j+j\wedge j=0$.
To see
this, let us only exemplify the calculation for the
$L_{ab}$-component of $j$, that is,
\begin{equation} \langle\d j+j\wedge j,L_{ab}\rangle\ =\ 0. \end{equation}
The others are verified in a similar manner.

To compute the projection of $\d j+j\wedge j$
onto the rotation generator $L_{ab}$, one first realizes that one needs
to consider
also the projections onto $P_a$ and $K_a$, respectively, which
follows upon inspecting the above commutation relations
of the Killing vectors. Then one finds
\begin{subequations}
\begin{equation}
 \langle \d j,L_{ab}\rangle\ =\ \d \langle j,L_{ab}\rangle
 \ =\ -\tfrac{2}{Y^4}\left[Y^i\d Y_i\wedge (\d X_aX_b-\d X_bX_a)+Y^2\d X_a\wedge\d X_b\right]\!.
\end{equation}
Similarly, one obtains ($\eta_{ac}\eta^{cb}={\delta_a}^b$)
\begin{equation}
\begin{aligned}
\langle j\wedge j,L_{ab}\rangle   \ &=\  2\eta^{cd}\langle j,L_{ac}\rangle
        \wedge\langle j, L_{db}\rangle+\langle j, P_{a}\rangle\wedge\langle j, K_{b}\rangle
        -\langle j, P_{b}\rangle\wedge\langle j, K_{a}\rangle\\
   \ &=\
\tfrac{2}{Y^4}\left[Y^i\d Y_i\wedge (\d X_aX_b-\d X_bX_a)+Y^2\d X_a\wedge\d X_b\right]\!,
\end{aligned}
\end{equation}
\end{subequations}
so that 
the combination of the two  expressions is indeed   zero.

Following the algorithm presented in Sec.~\ref{sec:gen}, one can then construct
two one-parameter families of flat currents and hence, an infinite number
of conserved non-local charges.
This familiar result  is, of course, 
 a consequence of
 $M={\rm AdS}_p\times S^{d-p}$ being a symmetric space.

\paragraph{Remark.}
In the following, we shall choose $r=0$ for notational simplicity, but 
all the relations  derived below are true also for a generic  space-time signature.

\section{\kern-10pt T-duality}

Subject of this section is the discussion of
certain integrable non-linear sigma models
and the derivation of their dual  cousins  by T-duality along certain
isometries. In particular, we will find the flat currents of T-dual models.

\subsection{\kern-10pt T-duality and flat currents for $S^2$}\label{sec:twosphere}

\paragraph{Setting.}
Before turning our attention to AdS geometries, we will consider
the two-sphere $S^2$ and perform T-duality along the compact
$U(1)$ isometry cycle. This example differs then from 
our main focus, i.e. the AdS$_p$ space, where  T-duality is
performed along non-compact isometric directions, but it serves to illustrate 
 in a simple setting a point that will apply  also to 
  AdS$_p$ geometries  -- that the  local symmetries of the
original model become hidden and are realized non-locally in the
T-dual theory.

The $S^2$ sigma model action reads
\begin{equation}\label{eq:actionS2}
S[\Phi,\Theta]\ =\ -\tfrac{1}{2}\int_\Sigma\left[
\d\Theta\wedge{*\d}\Theta+\sin^2(\Theta)\,\d\Phi\wedge{*\d}\Phi\right]\!.
\end{equation}
Following the procedure in Sec.~\ref{sec:general}, the Noether
currents are  found to be
\begin{equation}
 \begin{aligned}
  j_1 \ &=\  \tfrac{1}{\sqrt{2}}\left[\sin(\Phi)\,\d\Theta+
          \sin(\Theta)\cos(\Theta)\cos(\Phi)\,\d\Phi\right]\!,\\
  j_2 \ &=\  \tfrac{1}{\sqrt{2}}\left[\cos(\Phi)\,\d\Theta-
          \sin(\Theta)\cos(\Theta)\sin(\Phi)\,\d\Phi\right]\!,\\
  j_3 \ &=\  \tfrac{1}{\sqrt{2}}\sin^2(\Theta)\,\d\Phi.
\end{aligned}
\end{equation}
Besides being conserved, these currents are also flat
\begin{equation}
 \d j_A +  {f_A}^{BC}j_B\wedge j_C\ =\ 0,\qquad{\rm with}\qquad
 {f_A}^{BC}\ =\ -\tfrac{1}{\sqrt{2}}k_{AD}\epsilon^{DBC}
\end{equation}
for $A,B,\ldots=1,2,3$.
Here, $\epsilon^{ABC}$ is totally antisymmetric with $\epsilon^{123}=1$ and the
Cartan-Killing form $k$ is given by $k=(k_{AB})={\rm diag}(-1,-1,-1)$.
Therefore, we find two one-parameter families of flat currents\footnote{Here and
in the following we shall suppress the subscript `$\lambda\pm$' in \rf{las}
 and write $J$
instead of $J_{\lambda\pm}$.}
\begin{equation}\la{ab}
  J\ =\ a j+b\,{*j},\quad\mathrm{with}\quad a\ =\ 
  \textstyle{\frac12}[1\pm\cosh(\lambda)]\quad\mathrm{and}
   \quad b\ =\ \textstyle{\frac12}\sinh(\lambda),
\end{equation}
i.e.
\begin{equation}
 \d J_A + {f_A}^{BC} J_B\wedge J_C\ =\ 0,\qquad{\rm with}\qquad J_A\ =\ a j_A+b\,{*j_A},
\end{equation}
from which infinitely many non-local conserved charges may
be derived.

\paragraph{T-duality.}
We will now perform T-duality along the $\Phi$-direction. As
usual, this may be implemented by starting with the first-order action
\cite{bush,roc}
\begin{equation}\label{eq:roc}
S[\Theta,A,\tilde\Phi]\ =\
-\tfrac{1}{2}\int_\Sigma\left[
\d\Theta\wedge{*\d}\Theta+\sin^2(\Theta)A\wedge{*A}+2\,\tilde\Phi\,
\d A\right]\!,
\end{equation}
where the one-form $A$ is an Abelian gauge potential and the field $\tilde\Phi$
(to be later interpreted as 
T-dual to $\Phi$) plays  the role of a 
Lagrange multiplier for the field strength $F=\d A$.\footnote{In
principle, the gauge potential $A$ might have non-trivial holonomies
around non-contractible loops. This can be avoided by requiring
$\tilde\Phi$  to have the appropriate periodicity. Since our main
interest in this paper will be T-duality along
non-compact directions, we will not discuss this issue further 
(see \cite{roc}).} By
integrating out $\tilde\Phi$, we see that the
gauge potential $A$ is pure gauge
\begin{equation}
A\ =\ \d\Phi,
\end{equation} and upon substitution into Eq.~\eqref{eq:roc}, we recover the original action
\eqref{eq:actionS2}.
 On the other hand, the variation with respect to $A$ yields
\begin{equation}
A\ =\ \frac{1}{\sin^2(\Theta)}\,{*\d}\tilde\Phi,
\end{equation}
which implies the relation
\begin{equation} \la{pp}
\d\tilde\Phi\ =\ \sin^2(\Theta)\,{*\d\Phi}\end{equation}
between the original field
$\Phi$ and its T-dual $\tilde\Phi$.
 In terms
of $\tilde\Phi$, the T-dual action is given by
\begin{equation} \tilde S[\tilde\Phi,\Theta]\ =\ -\tfrac{1}{2}\int_\Sigma\left[
\d\Theta\wedge{*\d}\Theta+\frac{1}{\sin^2(\Theta)}\,\d\tilde\Phi\wedge{*\d}\tilde\Phi\right]\!.
\end{equation}

While the original  action was $SO(3)$   invariant, 
 the manifest symmetry  of the T-dual  action is simply    $U(1)$ shifts of $\tilde\Phi$.
Therefore, if we consider only the
Noether currents of the T-dual model, we will never be able to see
the full $SO(3)$ symmetry group which the T-dual model should also
admit, given the two sets of equations are  equivalent.    
This non-Abelian  symmetry group of the T-dual model, which we shall call 
 ``T-dual symmetry group''
and   denote by
$\widetilde{SO(3)}$,  will  be hidden and,  in addition, 
 realized non-locally.

\paragraph{T-dual symmetry group.}
To find this T-dual symmetry group, let us go back to
the flat currents of the original model. The T-duality transformation \rf{pp} 
cannot be directly  performed on the currents 
since they  depend  not only on $\d \Phi$ but also explicitly on the
coordinate $\Phi$.
  Note, however,  that flat currents
$\d J+J\wedge J=0$ are unique only up to a $G$-gauge transformation
($G=SO(3)$ in the present case)
\begin{equation} \la{gah}
J\ \mapsto\ J'\ =\ g^{-1}Jg+g^{-1}\d g,
\end{equation}
where  $g\in G$:  $J'$ is again flat.
Then there exists an element  $g:\Sigma\to SO(3)$ which transforms
the original currents into a new gauge-equivalent set of 
currents that depend  on $\Phi$  only through its derivatives.

To construct such  $g$, let us work in the fundamental
representation of the group $SO(3)$. The generators of $SO(3)$
obey
\begin{equation}
[T_A,T_B]\ =\ {f_{AB}}^CT_C,\qquad{\rm with}\qquad {f_{AB}}^C\ =\ -\tfrac{1}{\sqrt{2}}
\epsilon_{ABD}k^{DC},
\end{equation}
where $k=(k_{AB})={\rm diag}(-1,-1,-1)$ is as before; note that
$\epsilon_{123}=-1$. 
They  can be chosen as  
\begin{equation}
\begin{aligned}
T_1\ =\ \frac{1}{\sqrt{2}}\begin{pmatrix}
          0 & 0 & 0 \\ 0 & 0 & 1 \\ 0 & -1 & 0
        \end{pmatrix}\!,\qquad
T_2\ =\ \frac{1}{\sqrt{2}}\begin{pmatrix}
          0 & 0 & -1 \\ 0 & 0 & 0 \\ 1 & 0 & 0
        \end{pmatrix}\!,\\
T_3\ =\ \frac{1}{\sqrt{2}}\begin{pmatrix}
          0 & 1 & 0 \\ -1 & 0 & 0 \\ 0 & 0 & 0
        \end{pmatrix}\!.\kern2.5cm
\end{aligned}
\end{equation}
Then  the current $J=J_A T^A=J_A k^{AB}T_B$ takes the following
matrix form:
\begin{equation}
J\ =\ \frac{1}{\sqrt{2}}\begin{pmatrix}
                             0 & -J_3 & J_2\\
                             J_3 & 0 & -J_1\\
                            -J_2 & J_1 & 0
                            \end{pmatrix}\!.
\end{equation}
One can check that the choice of 
\begin{equation}
g\ =\ \begin{pmatrix}
        \cos(\Phi) & \sin(\Phi) & 0\\
            -\sin(\Phi) & \cos(\Phi) & 0\\
          0 & 0 & 1
         \end{pmatrix}
\end{equation}
yields the desired result:
\begin{equation}
 J'\ =\ g^{-1}Jg +g^{-1}\d g\ =\ \frac{1}{\sqrt{2}}\begin{pmatrix}
                             0 & -J'_3 & J'_2\\
                             J'_3 & 0 & -J'_1\\
                            -J'_2 & J'_1 & 0
                            \end{pmatrix}
\end{equation}
with (recall  that $a^2-a-b^2=0$, see \rf{ab})
\begin{equation}
 \begin{aligned}
 J'_1 \ &=\ \tfrac{1}{\sqrt{2}}\sin(\Theta)\cos(\Theta)\,(a\,\d\Phi+b\,{*\d\Phi}),\\
 J'_2  \ &=\  \tfrac{1}{\sqrt{2}}(a\,\d\Theta+b\,{*\d\Theta}),\\
 J'_3  \ &=\  \tfrac{1}{\sqrt{2}}\sin^2(\Theta)\,(a\,\d\Phi+b\,{*\d\Phi})-\sqrt{2}\,
  \d\Phi.
\end{aligned}
\end{equation}
Using \rf{pp},  we then find 
 the currents of the T-dual model: $\tilde J_A := J_A' (\Phi \mapsto \td \Phi)$, i.e.
\begin{equation}
 \begin{aligned}
 \tilde J_1  \ &=\  \tfrac{1}{\sqrt{2}}\cot(\Theta)\,(a\,{*\d\tilde\Phi}+b\,{\d\tilde\Phi}),\\
 \tilde J_2  \ &=\  \tfrac{1}{\sqrt{2}}(a\,\d\Theta+b\,{*\d\Theta}),\\
 \tilde J_3  \ &=\  \tfrac{1}{\sqrt{2}}(a\,{*\d\tilde\Phi}+b\,{\d\tilde\Phi})-
  \tfrac{\sqrt{2}}{\sin^2(\Theta)}{*\d\tilde\Phi}.
\end{aligned}
\end{equation}
They are again flat, $\d\tilde J+\tilde J\wedge\tilde J=0$, since
the relation $\d\tilde\Phi=\sin^2(\Theta){*\d\Phi}$ holds on-shell.

Proceeding as in Sec.~\ref{sec:general}, we get then infinitely
many conserved non-local charges. However, contrary to the
original model we  started with, the lowest order
charges\footnote{Here, we are considering the solution
$a=\frac12[1-\cosh(\lambda)]$ and $b=\frac12\sinh(\lambda)$ and
expand around $\lambda=0$.} are non-local and do not
correspond to Noether charges. In this sense, local charges are
mapped into non-local ones via T-duality. Conversely, the $U(1)$
Noether charge of the T-dual model is mapped into a non-local
charge in the original model. 

The AdS geometries discussed below
are quite different in this respect as they allow for a maximal
set of Noether charges in both the original and the T-dual model.
This means that both the original  sigma model action on the 
AdS$_p$ space and the one obtained after performing T-duality
along all the isometric directions, possess an $SO(p-1,1)$
symmetry group associated with the Noether charges.

\subsection{\kern-10pt T-duality and flat currents for AdS$_2$}

Next, let us consider the simplest AdS example:  AdS$_2$. 
The gauge 
 transformation of the flat current 
 constructed for this case   will turn out to be
the building block for all higher-dimensional AdS cases.

\paragraph{Setting.}
As before, we choose the conformally flat metric on AdS$_2$
\begin{equation}
S[X,Y]\ =\
-\tfrac{1}{2}\int_\Sigma\tfrac{1}{Y^2} \left(\d X\wedge{*\d X}+\d Y\wedge{*\d Y}\right)\!.
\end{equation}
The Noether currents have been derived in Sec.~\ref{sec:general},
and we repeat them here:
\begin{equation}
 \begin{aligned} \la{kar}
   j_1  \ &=\  -\tfrac{1}{2\sqrt{2}Y^2}\left\{\left[1+(X^2-Y^2)\right]\d X+2XY\d Y\right\}\!,\\
   j_2  \ &=\  \tfrac{1}{2\sqrt{2}Y^2}\left\{\left[1-(X^2-Y^2)\right]\d X-2XY\d Y\right\}\!,\\
   j_3  \ &=\  -\tfrac{1}{\sqrt{2}Y^2}\left(X\d X+Y\d Y\right)\!.\\
\end{aligned}
\end{equation}
Here, we have taken certain linear combinations of the translation and special
conformal currents and also introduced different normalization constants.
As a result,   these   currents are flat, i.e.~they obey
\begin{equation}
\d j_A+{f_A}^{BC}j_B\wedge j_C\ =\ 0,\qquad{\rm with}\qquad
   {f_A}^{BC}\ =\ \tfrac{1}{\sqrt{2}}k_{AD}\epsilon^{DBC},
\end{equation}
where the Cartan-Killing form $k$ is given by $k=(k_{AB})={\rm diag}(-1,1,1)$.
Again, we may introduce a family of  flat currents according to
\begin{equation}\la{karr}
J\ =\ a\, j+b\,{*j}.
\end{equation}

\paragraph{T-duality.}
Performing T-duality along the $X$-direction by repeating the steps that led to the
T-dual action in the $S^2$ example, we find
\begin{equation}\la{jk}
\d\tilde X\ =\ \tfrac{1}{Y^2} {*\d X}
\end{equation}
and hence,
\begin{equation}
\tilde S[\tilde X,Y]\ =\
-\tfrac{1}{2}\int_\Sigma
\left(Y^2\d\tilde X\wedge{*\d\tilde X}+\tfrac{1}{Y^2} \d Y\wedge{*\d Y}\right)\!.
\end{equation}
This  is again an AdS$_2$ sigma model. Getting back
after T-duality the space one has started with is a special feature of AdS
geometries (cf.  $S^2$ example).\footnote{In fact, any geometry with a metric of the form
 $$g\ =\ f(Y)\,\d Y\otimes\d Y+h(Y)\,\d X\otimes\d X$$
goes back into itself after performing a T-duality
along isometric direction $X$  provided 
and $Y=Y(\tilde Y)$ with $[Y'(\tilde Y)]^2=f(\tilde Y)/f(Y(\tilde Y))$ and
$h(\tilde Y)=[h(Y(\tilde Y))]^{-1}$. A  simple example is 
$f=1$, $h(Y) =\exp(Y)$ and
$Y=-\tilde Y$.} 

Therefore, the T-dual model  also exhibits an $SO(2,1)$
isometry group. Upon constructing the T-dual Noether currents (in the
T-dual coordinates), one may derive flat
currents in the T-dual model which induce infinitely many conserved non-local
charges. As in the original model, the lowest order charges are the Noether
charges. This time,  these are, of course, the Noether charges for the T-dual model
in the T-dual coordinates.

\paragraph{T-dual symmetry group.}
Let us now discuss a similar procedure as in  the two-sphere example, that is,
let us show that there exists a transformation mediated by 
$g:\Sigma\to SO(2,1)$ which maps 
the original currents into a new set of gauge equivalent
currents that  depend on $X$  only through its derivatives.

As before, we shall work in  the fundamental representation. The generators obey
\begin{equation}
[T_A,T_B]\ =\ {f_{AB}}^CT_C,\qquad{\rm with}\qquad {f_{AB}}^C\ =\ \tfrac{1}{\sqrt{2}}
\epsilon_{ABD}k^{DC},
\end{equation}
where $k=(k_{AB})={\rm diag}(-1,1,1)$ is the Cartan-Killing form; 
note that $\epsilon_{123}=-1$.

The generators  can be chosen as 
\begin{equation}
 \begin{aligned}
T_1\ =\ \frac{1}{\sqrt{2}}\begin{pmatrix}
          0 & 1 & 0 \\ -1 & 0 & 0 \\ 0 & 0 & 0
        \end{pmatrix}\!,\qquad
T_2\ =\ \frac{1}{\sqrt{2}}\begin{pmatrix}
          0 & 0 & -1 \\ 0 & 0 & 0 \\ -1 & 0 & 0
        \end{pmatrix}\!,\\
T_3\ =\ \frac{1}{\sqrt{2}}\begin{pmatrix}
          0 & 0 & 0 \\ 0 & 0 & -1 \\ 0 & -1 & 0
        \end{pmatrix}\!.\kern2.5cm
\end{aligned}
\end{equation}
so that  $J=J_A T^A=J_A k^{AB}T_B$ is given by
\begin{equation} J\ =\ \frac{1}{\sqrt{2}}\begin{pmatrix}
                             0 & -J_1 & -J_2\\
                             J_1 & 0 & -J_3\\
                            -J_2 & -J_3 & 0
                            \end{pmatrix}\!.
\end{equation}

Then a short calculation reveals that 
\begin{equation}
g\ =\ \begin{pmatrix}
       X & 1 & -X\\
       1-\frac12 X^2 & -X & \frac12 X^2\\
       \frac12 X^2 & X & -1-\frac12 X^2
      \end{pmatrix}\ \in\ SO(2,1)
\end{equation}
gauges away the explicit  $X$-dependence of the currents \rf{kar} and \rf{karr}.
The components of $J'=g^{-1}Jg+g^{-1}\d g$  are found to be 
 \begin{equation} \la{klo}
  \begin{aligned}
J'_1 \ &=\  \tfrac{1}{2\sqrt{2}Y^2}(1-Y^2)(a\, \d X+b\,{*\d X})+\sqrt{2}\,\d X,\\
J'_2 \ &=\  \tfrac{1}{\sqrt{2}Y}(a\,\d Y+b\,{*\d Y}),\\
J'_3 \ &=\ -\tfrac{1}{2\sqrt{2}Y^2}(1+Y^2)(a\,\d X+b\,{*\d X})+\sqrt{2}\,\d X.\\
\end{aligned}
\end{equation}
 The T-dual currents 
 \begin{equation} \la{tj}  \tilde J_A (Y, \td X)\ :=\ J'_A(Y, X(\tilde X))   
 \end{equation}
 are then  given by
\begin{equation}\label{eq:tdualcurrents}
 \begin{aligned}
 \tilde J_1  \ &=\  \tfrac{1}{2\sqrt{2}}(1-Y^2)(a\,{*\d\tilde X}+b\,{\d\tilde X})+\sqrt{2}Y^2
    \,{*\d\tilde X},\\
 \tilde J_2  \ &=\  \tfrac{1}{\sqrt{2}Y}(a\,\d Y+b\,{*\d Y}),\\
 \tilde J_3 \ &=\  -\tfrac{1}{2\sqrt{2}}(1+Y^2)(a\,{*\d\tilde X}+b\,{\d\tilde X})+\sqrt{2}Y^2\,
    {*\d\tilde X}, 
\end{aligned}
\end{equation}
and  are  again flat (recall that  $a^2-a-b^2=0$).

Note that the lowest order charges obtained by
expanding the path-ordered exponential \eqref{pathorder} around
$\lambda=0$ are not the Noether charges for the T-dual model. The
latter follow from the T-dual flat currents $\hat J=\hat a\hat j+\hat b\,{*\hat j}$, where $\hat j$ is now
given by
\begin{equation}
 \begin{aligned}
 \hat j_1 \ &=\  -\tfrac{1}{2\sqrt{2}}\big\{\big[Y^2-(1-\tilde X^2Y^2)\big]\d\tilde X
                 -\tfrac{2\tilde X}{Y}\d Y\big\},\\
 \hat j_2 \ &=\  \tfrac{1}{2\sqrt{2}}\big\{\big[Y^2+
                    (1-\tilde X^2Y^2)\big]\d\tilde X+\tfrac{2\tilde X}{Y}\d Y\big\},\\
 \hat j_3 \ &=\  -\tfrac{1}{\sqrt{2}}\big(Y^2\tilde X\d\tilde X-\tfrac{1}{Y}\d Y\big)
\end{aligned}
\end{equation}
and $\hat a^2-\hat a-\hat b^2=0$, with
\begin{equation}\hat a\ =\ \tfrac{1}{2}[1\pm\cosh(\hat\lambda)]\qquad\mbox{and}
\qquad \hat b\ =\ \tfrac{1}{2}\sinh(\hat\lambda).
\end{equation}
By the above reasoning, the $\tilde X$-dependence of $\hat J$ can be gauged away to get
 \begin{equation}\label{eq:tdualnoether}
  \begin{aligned}
 \hat J'_1  \ &=\  -\tfrac{1}{2\sqrt{2}}(1-Y^2)(\hat a\,\d\tilde X+\hat b\,{*\d\tilde X})+\sqrt{2}\,\d\tilde X,\\
 \hat J'_2  \ &=\  \tfrac{1}{\sqrt{2}Y}(\hat a\,\d Y+\hat b\,{*\d Y}),\\
 \hat J'_3  \ &=\  -\tfrac{1}{2\sqrt{2}}(1+Y^2)(\hat a\,\d\tilde X+\hat b\,{*\d\tilde X})+\sqrt{2}\,\d\tilde X.
\end{aligned}
\end{equation}
In this sense, we  get the following relation:
\begin{center}
\begin{tabular}{ccc}
     \underline{\bf Original model} &  & \underline{\bf T-dual model}\\[5pt]
  local $SO(2,1)$  &$\Longrightarrow$ & non-local $\widetilde{SO(2,1)},$\\
  non-local $\widetilde{SO(2,1)}$  &$\Longleftarrow$ & local $SO(2,1).$\\
\end{tabular}
\end{center}

\subsection{T-duality and flat currents for AdS$_5$}

Let us now turn to  higher-dimensional AdS geometries. For concreteness,
we shall stick to AdS$_5$ but the derivations presented below can
straightforwardly be extended to any dimension by means of a
recursive procedure. As already indicated, the transformation for the
AdS$_2$ space derived above will form a basic building block.

\paragraph{Setting.}
As before, we shall start with  the   Euclidean ($r=0$) 
AdS  metric in the conformally flat  form\footnote{The indices $a,b=1,\ldots,4$ 
here are contracted with $\delta_{ab}$.}
\begin{equation}
 g\ =\ \tfrac{1}{Y^2}(\,\d X_a\otimes\d X_a+\d Y\otimes\d Y),
\end{equation}
The Noether currents for the corresponding sigma model 
 are
\begin{equation}\label{eq:noetherads5}
 \begin{aligned}
   j_1  \ &=\  -\tfrac{1}{2\sqrt{2}Y^2}\left\{\left[1+X_1^2-(X_{(1)}\cdot X_{(1)}+Y^2)\right]\d X_1
              +2X_1(X_{(1)}\cdot\d X_{(1)}+Y\d Y)\right\}\!,\\
   j_2  \ &=\  \tfrac{1}{2\sqrt{2}Y^2}\left\{\left[1-X_1^2+(X_{(1)}\cdot X_{(1)}+Y^2)\right]\d X_1
              -2X_1(X_{(1)}\cdot\d X_{(1)}+Y\d Y)\right\}\!,\\
   j_3  \ &=\  -\tfrac{1}{\sqrt{2}Y^2}\left(X_1\d X_1+X_{(1)}\cdot\d X_{(1)}+Y\d Y\right)\!,\\[6pt]
   j_4  \ &=\  -\tfrac{1}{2\sqrt{2}Y^2}\left\{\left[1+X_2^2-(X_{(2)}\cdot X_{(2)}+Y^2)\right]\d X_2
              +2X_2(X_{(2)}\cdot\d X_{(2)}+Y\d Y)\right\}\!,\\
   j_5  \ &=\  \tfrac{1}{2\sqrt{2}Y^2}\left\{\left[1-X_2^2+(X_{(2)}\cdot X_{(2)}+Y^2)\right]\d X_2
              -2X_2(X_{(2)}\cdot\d X_{(2)}+Y\d Y)\right\}\!,\\
   j_6  \ &=\  -\tfrac{1}{\sqrt{2}Y^2}\left(X_1\d X_2-X_2\d X_1\right)\!,\\[6pt]
   j_7  \ &=\  -\tfrac{1}{2\sqrt{2}Y^2}\left\{\left[1+X_3^2-(X_{(3)}\cdot X_{(3)}+Y^2)\right]\d X_3
              +2X_3(X_{(3)}\cdot\d X_{(3)}+Y\d Y)\right\}\!,\\
   j_8  \ &=\  \tfrac{1}{2\sqrt{2}Y^2}\left\{\left[1-X_3^2+(X_{(3)}\cdot X_{(3)}+Y^2)\right]\d X_3
              -2X_3(X_{(3)}\cdot\d X_{(3)}+Y\d Y)\right\}\!,\\
   j_9  \ &=\  -\tfrac{1}{\sqrt{2}Y^2}\left(X_1\d X_3-X_3\d X_1\right)\!,\\
   j_{10}  \ &=\  -\tfrac{1}{\sqrt{2}Y^2}\left(X_2\d X_3-X_3\d X_2\right)\!,\\[6pt]
  j_{11}  \ &=\  -\tfrac{1}{2\sqrt{2}Y^2}\left\{\left[1+X_4^2-(X_{(4)}\cdot X_{(4)}+Y^2)\right]\d X_4
              +2X_4(X_{(4)}\cdot\d X_{(4)}+Y\d Y)\right\}\!,\\
   j_{12}  \ &=\  \tfrac{1}{2\sqrt{2}Y^2}\left\{\left[1-X_4^2+(X_{(4)}\cdot X_{(4)}+Y^2)\right]\d X_4
              -2X_4(X_{(4)}\cdot\d X_{(4)}+Y\d Y)\right\}\!,\\
   j_{13}  \ &=\  -\tfrac{1}{\sqrt{2}Y^2}\left(X_1\d X_4-X_4\d X_1\right)\!,\\
   j_{14}  \ &=\  -\tfrac{1}{\sqrt{2}Y^2}\left(X_2\d X_4-X_4\d X_2\right)\!,\\
   j_{15}  \ &=\  -\tfrac{1}{\sqrt{2}Y^2}\left(X_3\d X_4-X_4\d X_3\right)\!,
\end{aligned}
\end{equation}
where we have introduced the abbreviation
\begin{equation}
  X_{(a)}\ :=\ (\cdots,X_{a-1},X_{a+1},\cdots).
\end{equation}
We shall also use $X_{(0)}:=(X_1,\ldots,X_4)$, in the sequel.
The `$\cdot$' refers to the (Euclidean) scalar product, e.g.
$X_{(1)}\cdot X_{(1)}=X_2^2+X_3^2+X_4^2$, etc.
Note that we have again taken certain linear combinations of the translation
and special conformal currents. The AdS$_2$ case can be recovered for 
 $X_{2,3,4}=0$, AdS$_3$ 
for $X_{3,4}=0$  and AdS$_4$ for $X_4=0$, respectively. 
 In addition,
the above currents have been grouped
according to their appearance: $j_{1,2,3}$ are the AdS$_2$ currents,
$j_{1,\ldots,6}$ the AdS$_3$ currents, $j_{1,\ldots,10}$ the AdS$_4$ currents
and  $j_{1,\ldots,15}$ the AdS$_5$ currents (after putting to zero the
appropriate $X_a$-coordinates).

These currents are flat
\begin{equation}
 \d j_A+{f_A}^{BC}j_B\wedge j_C\ =\ 0,\qquad{\rm for}\qquad A,B,\ldots=1,\ldots,15,
\end{equation}
where ${f_{AB}}^C$ are the structure constants of
$\mathfrak{so}(5,1)$. Again, we may introduce the one-parameter
family of flat currents $J= aj+b\,{*j}$, with  $a^2-a-b^2=0$.

\paragraph{T-duality.}
Our main  observation 
 is that, as in the AdS$_2$ example, one is able to gauge away the
$X_a$-dependence by means of a field dependent gauge transformation.
This makes it  possible to  perform T-duality along the $X_a$-directions
in a straightforward way.

As before, we work in the fundamental representation of $SO(5,1)$.
Let us denote the generators by $T_A$. The current
$J=J_A T^A=J_A k^{AB}T_B$, with $J_A=aj_A+b\,{*j_A}$ and $j_A$ given in \eqref{eq:noetherads5},
might be represented as\footnote{Recall that the generators in
the fundamental representation are given by
$$(t_{\alpha\beta})^{\gamma\delta}\ =\ {\delta_\alpha}^\gamma{\delta_\beta}^\delta
-{\delta_\beta}^\gamma{\delta_\alpha}^\delta,\qquad{\rm with}\qquad
\alpha,\beta,\ldots\ =\ 1,\ldots,6.$$
Upon relabeling the set $\{t_{\alpha\beta}\}\mapsto \{T_A\}$, one obtains the present
choice of parametrization.}
\begin{equation}\label{eq:gencur}
 J\ =\ \frac{1}{\sqrt{2}}\begin{pmatrix}
                            0       & -J_{15} & -J_{14} & -J_{13} & -J_{11} & -J_{12}\\
                            J_{15}  & 0       & -J_{10} & -J_9    & -J_7    & -J_8\\
                            J_{14}  & J_{10}  & 0       & -J_6    & -J_4    & -J_5\\
                            J_{13}  & J_9     & J_6     & 0       & -J_1    & -J_2\\
                            J_{11}  & J_7     & J_4     & J_1     & 0       & -J_3\\
                            -J_{12} & -J_8    & -J_5    & -J_2    & -J_3    & 0
                         \end{pmatrix}.
\end{equation}
Here, the Cartan-Killing form is given by
\begin{equation}
 (k_{AB})\ =\ {\rm diag}(-1,1,1,-1,1,-1,-1,1,-1,-1,-1,1,-1,-1,-1).
\end{equation}
Notice that the lower right $3\times3$-block in Eq.~\eqref{eq:gencur} represents the
AdS$_2$ case, the $4\times4$-block the AdS$_3$ case, etc.

The key idea is to gauge away the explicit
dependence on the $X_a$-coordinates recursively, i.e.~first $X_1$, then $X_2$, etc.
Following this algorithm, one finds that the 
required gauge transformation matrix is 
\begin{subequations}
\begin{equation}
  g\ =\ g_1g_2g_3g_4,
\end{equation}
with
\begin{equation}
  g_1\ =\ \begin{pmatrix}
             1 & 0 & 0 & 0 & 0 & 0\\
             0 & 1 & 0 & 0 & 0 & 0\\
             0 & 0 & 1 & 0 & 0 & 0\\
             0 & 0 & 0 & X_1 & 1 & -X_1\\
             0 & 0 & 0 & 1-\frac12 X_1^2 & -X_1 & \frac12 X_1^2\\
             0 & 0 & 0 & \frac12 X_1^2 & X_1 & -1-\frac12 X_1^2
          \end{pmatrix},
\end{equation}
\begin{equation}
  g_2\ =\ \begin{pmatrix}
             1 & 0 & 0 & 0 & 0 & 0\\
             0 & 1 & 0 & 0 & 0 & 0\\
             0 & 0 & X_2 & 1 & -\im X_2 & 0\\
             0 & 0 & 1-\frac12 X_2^2 & -X_2 & \frac{\im}{2} X_2^2 & 0\\
               0 & 0 & 0 & 0 & 0 & -\im\\
             0 & 0 & -\frac12 X_2^2 & -X_2 & \im+\frac{\im}{2} X_2^2 & 0
          \end{pmatrix},
\end{equation}
\begin{equation}
  g_3\ =\ \begin{pmatrix}
             1 & 0 & 0 & 0 & 0 & 0\\
             0 & X_3 & 1 & -\im X_3 & 0 & 0\\
             0 & 1-\frac12 X_3^2 & -X_3 & \frac{\im}{2} X_3^2 & 0 & 0\\
             0 & 0 & 0 & 0 & 1 & 0\\
             0 & \frac{\im}{2} X_3^2 & \im X_3 & 1+\frac{1}{2} X_3^2 & 0 & 0\\
             0 & 0 & 0 & 0 & 0 & 1
          \end{pmatrix},
\end{equation}
\begin{equation}
  g_4\ =\ \begin{pmatrix}
             X_4 & 1 & -\im X_4 & 0 & 0 & 0\\
             1-\frac12 X_4^2 & -X_4 & \frac{\im}{2} X_3^2 & 0 & 0 & 0\\
             0 & 0 & 0 & 1 & 0 & 0\\
             \frac{\im}{2} X_4^2 & \im X_4 & 1+\frac{1}{2} X_4^2 & 0 & 0 & 0\\
             0 & 0 & 0 & 0 & 1 & 0\\
             0 & 0 & 0 & 0 & 0 & 1
          \end{pmatrix},
\end{equation}
i.e.
\begin{equation}
 g\ =\ \begin{pmatrix}
 X_4 & 1 & -\im X_4 & 0 & 0 & 0 \\
 X_3 & 0 & -\im X_3 & 1 & 0 & 0 \\
 X_2 & 0 & -\im X_2 & 0 & 1 & 0 \\
 X_1 & 0 & -\im X_1 & 0 & 0 & -\im \\
 1-\frac12 X_{(0)}\cdot X_{(0)} & -X_4 & \frac{\im}{2}
   X_{(0)}\cdot X_{(0)} & -X_3 & -X_2 & \im X_1 \\
 \frac{1}{2} X_{(0)}\cdot X_{(0)} & X_4 & -\im-\frac{\im}{2}X_{(0)}\cdot X_{(0)}
   & X_3 & X_2 & -\im X_1
\end{pmatrix}.
\end{equation}
\end{subequations}
As one can check,
$g_1\in SO(5,1)$ while $g_{2,3,4}$ are elements of  the complexified gauge group
$SO_\IC(5,1)$ (here, $\im:=\sqrt{-1}$). Hence, $g\in SO_\IC(5,1)$.

Let us stress that 
the appearance of the complexified
gauge algebra $\mathfrak{so}_\IC(5,1)$ and gauge group $SO_\IC(5,1)$ is
merely a consequence of the fact that we have chosen to start 
with the  Euclidean AdS$_5$ space.
 Indeed, if one instead starts with 
Minkowski (or Kleinian, i.e.~split signature) AdS$_5$ space, 
all the four $g_m$-matrices  ($m=1,\ldots,4$)  live in
$SO(4,2)$ (or $SO(3,3)$); this can also be seen by making 
a suitable Wick rotation.\footnote{\label{foot:embedding} More concretely, 
let $Z_{1,\ldots, 6}$ be coordinates on $\IR^{5,1}$. 
Euclidean AdS$_5$ can be viewed as the hyper-surface 
$Z_1^2+\cdots +Z_5^2-Z_6^2=-1$ in $\IR^{5,1}$. 
The relation between embedding and
Poincar\'e coordinates, suitable for our present purposes,
is then $Z_2=X_2/Y$, $Z_3=X_1/Y$, $Z_4=X_3/Y$, $Z_5=X_4/Y$,
 $Z_6+Z_1=1/Y$ and $Z_6-Z_1=[Y^2+X_{(0)}\cdot X_{(0)}]/Y^2$.
To get real currents in Minkowski
AdS$_5$, one sends $X_1$ to $\im X_1$ and finally does a reparametrization
of the embedding coordinates according to $Z_3\leftrightarrow Z_4$. 
Alternatively, one may directly start with the currents \eqref{eq:noetherads5}
in Minkowski signature, go through the procedure described in the main text and
verify explicitly that the $g_m$-matrices are real.}

It remains to write down  the gauge transformed currents,
\begin{equation}
  J'\ =\ g^{-1} J g +g^{-1}\d g.
\end{equation}
Their non-vanishing components read as
\begin{equation}\la{jjj}
 \begin{aligned}
   J'_4\ &=\ \tfrac{\im}{2\sqrt{2}Y^2}(1+Y^2)(a\,\d X_2+b\,{*\d}X_2)-\sqrt{2}\im\,\d X_2,\\
   J'_5\ &=\ \tfrac{1}{2\sqrt{2}Y^2}(1+Y^2)(a\,\d X_1+b\,{*\d}X_1)-\sqrt{2}\,\d X_1,\\
   J'_6\ &=\ \tfrac{\im}{2\sqrt{2}Y^2}(1+Y^2)(a\,\d X_3+b\,{*\d}X_3)-\sqrt{2}\im\,\d X_3,\\
   J'_{10}\ &=\ \tfrac{\im}{2\sqrt{2}Y^2}(1+Y^2)(a\,\d X_4+b\,{*\d} X_4)+\sqrt{2}\im\,\d X_4,\\
   J'_{11}\ &=\ \tfrac{1}{2\sqrt{2}Y^2}(1-Y^2)(a\,\d X_2+b\,{*\d} X_2)+\sqrt{2}\,\d X_2,\\
   J'_{12}\ &=\ \tfrac{\im}{2\sqrt{2}Y^2}(1-Y^2)(a\,\d X_1+b\,{*\d} X_1)-\sqrt{2}\im\,\d X_1,\\
   J'_{13}\ &=\ \tfrac{1}{2\sqrt{2}Y^2}(1-Y^2)(a\,\d X_3+b\,{*\d} X_3)+\sqrt{2}\,\d X_3,\\
   J'_{14}\ &=\ \tfrac{\im}{\sqrt{2}Y}(a\,\d Y+b\,{*\d Y}),\\
   J'_{15}\ &=\ \tfrac{1}{2\sqrt{2}Y^2}(1-Y^2)(a\,\d X_4+b\,{*\d} X_4)+\sqrt{2}\,\d X_4.
 \end{aligned}
\end{equation}
Now the T-duality can easily be implemented by using\footnote{Notice that we work with a Minkowski world-sheet. If one instead considers a Euclidean
world-sheet, one has $\d X_a =\im Y^2{*\d}\tilde X_a$. Then the T-dual currents will be
real, i.e.~$\mathfrak{so}(5,1)$-valued, if we choose the relation between the Poincar\'e and
embedding coordinates as in footnote \ref{foot:embedding}.}
\begin{equation}\label{eq:TdualityAdS5}
 \d X_a\ =\ Y^2{*\d}\tilde X_a,
\end{equation}
which yields the T-dual metric
\begin{equation}
 \tilde g\ =\ Y^2\d\tilde X_a\otimes\d\tilde X_a+\tfrac{1}{Y^2} \d Y\otimes\d Y.
\end{equation}
This leads to 
the  following expressions for the  T-dual  currents in \rf{tj}:  
$\tilde J (Y, \td X_a) =  J(Y, X_a(\tilde X_b)) = \tilde J_A T^A$   with   
\begin{equation}
 \begin{aligned}
   \tilde J_4\ &=\ \tfrac{\im}{2\sqrt{2}}(1+Y^2)(a\,{*\d}
        \tilde X_2+b\,\d\tilde X_2)-\sqrt{2}\im Y^2{*\d} \tilde X_2,\\
   \tilde J_5\ &=\
     \tfrac{1}{2\sqrt{2}}(1+Y^2)(a\,{*\d} \tilde X_1+b\,\d\tilde X_1)-\sqrt{2} Y^2{*\d} \tilde X_1,\\
   \tilde J_6\ &=\ \tfrac{\im}{2\sqrt{2}}(1+Y^2)(a\,{*\d}
       \tilde X_3+b\,\d\tilde X_3)-\sqrt{2}\im Y^2{*\d} \tilde X_3,\\
   \tilde J_{10}\ &=\ \tfrac{\im}{2\sqrt{2}}(1+Y^2)
                       (a\,{*\d} \tilde X_4+b\,\d \tilde X_4)+\sqrt{2}\im Y^2{*\d} \tilde X_4,\\
   \tilde J_{11}\ &=\ \tfrac{1}{2\sqrt{2}}(1-Y^2)(a\,{*\d} \tilde X_2+b\,\d \tilde X_2)+
         \sqrt{2} Y^2{*\d} \tilde X_2,\\
   \tilde J_{12}\ &=\ \tfrac{\im}{2\sqrt{2}}(1-Y^2)
                   (a\,{*\d} \tilde X_1+b\,\d \tilde X_1)-\sqrt{2}\im Y^2{*\d} \tilde X_1,\\
   \tilde J_{13}\ &=\ \tfrac{1}{2\sqrt{2}}(1-Y^2)(a\,{*\d} \tilde X_3+b\,\d \tilde X_3)+
       \sqrt{2} Y^2{*\d} \tilde X_3,\\
   \tilde J_{14}\ &=\ \tfrac{\im}{\sqrt{2}Y}(a\,\d Y+b\,{*\d} Y),\\
   \tilde J_{15}\ &=\ \tfrac{1}{2\sqrt{2}}(1-Y^2)(a\,{*\d} \tilde X_4+b\,\d \tilde X_4)+
           \sqrt{2} Y^2{*\d} \tilde X_4.
 \end{aligned}
\end{equation}
As in the AdS$_2$ case, these currents are flat since
 \eqref{eq:TdualityAdS5} holds on-shell.

\paragraph{Remarks.}
The above derivation can be extended to sigma models on 
any AdS$_p$ space; one simply adds the additional currents to the above
set and performs successive gauge transformations.

It should be noted  that one may gauge away say only the $X_1$  coordinate dependence 
to perform the 
T-duality only along the $X_1$-direction. In this case, the isometry
group of the T-dual model is  only a subgroup of $SO(5,1)$. Nevertheless,
the T-dual currents lead to the  full $SO(5,1)$ symmetry,\footnote{More precisely
it is $SO_\IC(5,1)$, since here we are using the Euclidean signature.} which we
may  denote, as in the  AdS$_2$  case, by 
$\widetilde{SO(5,1)}$. Hence, even though the isometry group
of  the T-dual space is smaller, the T-dual model always  has   a ``hidden'' 
$\widetilde{SO(5,1)}$ symmetry.
This is in the same spirit as in 
the $S^2$ example discussed in Sec.~\ref{sec:twosphere}
and shows again that,  under the  T-duality, local (Noether)
charges of the original model are mapped to non-local charges 
of  the T-dual model and vice versa. However, unlike the $S^2$ example,
after T-duality along all $X_a$-directions, one recovers
the same maximal symmetry group, now being generated by the
Noether charges of the
T-dual model.

\bs \bs

\noindent{\bf Acknowledgements.}~We  
are grateful to F. Alday, G. Arutyunov, J. de Boer, 
 D. Gross, A. Gorsky and R. Roiban
for useful discussions. We 
 acknowledge  the support of the  EU under the MRTN contract MRTN--CT--2004--005104.
 A.A.T. was also supported by the  INTAS 03--51--6346 grant  and the RS Wolfson award. 
M.W. was supported in part 
 by the STFC under the rolling grant PP/D0744X/1.
Part of this work was done  during the periods when A.A.T. and M.W. were 
participants of the  program ``Strong Fields, Integrability and Strings'' at the Isaac
Newton Institute in Cambridge, U.K.  
and we thank the organizers for their invitation and hospitality.

\bs\bs

\end{document}